\documentclass[12pt]{article}
\usepackage[dvips]{graphicx}

\usepackage{epsfig,amsmath,amssymb,verbatim,mathrsfs}

\def\beq{\begin{eqnarray}}
\def\eeq{\end{eqnarray}}
\def\bea{\begin{eqnarray}}
\def\eea{\end{eqnarray}}

\newcommand{\D}{\text{\tiny D}}

\newcommand{\ri}{\text{\tiny R}}
\newcommand{\li}{\text{\tiny L}}
\newcommand{\nl}{\text{\tiny N}}

\newcommand{\be}{\begin{equation}}
\newcommand{\ee}{\end{equation}}

\begin{document}
\begin{titlepage}
\noindent

\setlength{\baselineskip}{0.2in}
\vspace{10cm}
\flushright{November 2010}

\vspace{1cm}

\begin{center}
  \begin{Large}
    \begin{bf}
Light Neutrinos from a Mini-Seesaw Mechanism in Warped Space
     \end{bf}
  \end{Large}
\end{center}
\vspace{0.2cm}

\begin{center}

\begin{large}
Kristian L. McDonald\\
\end{large}
\vspace{0.8cm}
  \begin{it}
Max-Planck-Institut f\"ur Kernphysik,\\
 Postfach 10 39 80, 69029 Heidelberg, Germany.\vspace{0.3cm}\\
Email: kristian.mcdonald@mpi-hd.mpg.de
\end{it}
\vspace{0.5cm}

\end{center}


\begin{abstract}
The seesaw mechanism provides a simple explanation for
the lightness of the known neutrinos. Under the
standard assumption of a weak scale Dirac mass and a heavy sterile
Majorana scale the neutrino
 mass is naturally suppressed below the weak scale. However, Nature may
 employ Dirac and Majorana scales that are much less than typically
 assumed, possibly
 even far below the weak scale. In this
 case the seesaw mechanism alone would not completely explain the lightness of
 the neutrinos. In this work we consider a
 warped framework
 that realizes this possibility by combining naturally suppressed Dirac
 and Majorana scales together in a mini-seesaw mechanism to generate
 light neutrino masses. Via the AdS/CFT correspondence the
 model is dual to a 4D theory with a hidden strongly coupled sector
 containing light
 composite right-handed neutrinos.
\end{abstract}

\vspace{1cm}

\end{titlepage}

\setcounter{page}{1}


\vfill\eject


\section{Introduction}
The confirmed discovery of neutrino mass has provided
 further information about the flavor structure of Nature. 
Despite the riches afforded by these developments we
 still do not know the underlying mechanism responsible for neutrino
 mass generation (for nice reviews see~\cite{Nu_review}). The
 seesaw mechanism provides a simple explanation for
the lightness of the known neutrinos~\cite{seesaw}. In the standard seesaw picture one assumes
$M_R\gg m_{\D}\sim m_W$, where $m_{D}$ 
 is the neutrino Dirac mass, $M_R$ is the singlet Majorana mass and
 $m_W$ is the $W$ boson mass. The resulting mass eigenstates include a
 ``mostly active'' light neutrino with mass $m_\nu\sim
 m_{\D}^2/M_R$ and a ``mostly sterile'' heavy neutrino with mass $\sim
 M_R$.  The suppressed mass of the lightest state nicely explains the
 existence of light neutrinos that interact weakly with standard model (SM)
 leptons. The existence of a large mass scale $M_R$ also marries well
 with our expectations for gauge unification in the ultraviolet (UV)
 completion of the SM.

Beyond expectations of naturalness, there
 is perhaps no particular reason to expect
 the Dirac mass to be of order $\sim m_W$. It is at
 least reasonable to consider Dirac masses in the range
 $m_e\le m_{\D} \le m_t$, consistent with those observed in the
 charged fermion sector~\cite{deGouvea:2005er}. Unfortunately, even at
 the light end of this
 range, values of $m_\nu\sim0.1$~eV require $M_R\sim 10$~TeV and it is
 difficult to probe the Majorana scale directly; a
 situation that becomes increasingly hopeless as $m_{\D}$ increases. From
 a naturalness point of view~\cite{'tHooft:1979bh}, arbitrarily light Dirac mass scales are
technically natural due to the restoration of a chiral symmetry in the
limit $m_{\D}\rightarrow 0$ with $M_R=0$. Indeed, even arbitrarily light Majorana mass
scales are technically natural as $M_R\rightarrow0$ restores lepton
number symmetry. However, despite being technically natural small values typically
require small couplings and one hopes that the discovery of neutrino
mass reveals more than a simple preference for tiny couplings in Nature.

An interesting exception to the standard seesaw picture arises if
the right-handed neutrinos are composite objects of a strongly
coupled hidden sector~\cite{ArkaniHamed:1998pf,Okui:2004xn,Grossman:2008xb}. The resulting neutrino mass
scales can be suppressed by powers of $(\Lambda_{hid}/\Lambda)$ and may be much lighter than typically assumed ($\Lambda_{hid}$ is the hidden confinement scale and $\Lambda$ is the
cutoff). A related
scenario is that with ``late-time neutrino masses'' wherein flavor
symmetries ensure massless neutrinos until relatively low energies,
after which symmetry breaking induces neutrino
mass~\cite{Chacko:2003dt}.  

In this work we develop an
approach to the generation of SM neutrino
mass that is inspired by the notion of light, composite right-handed
neutrinos~\cite{ArkaniHamed:1998pf}. The approach realizes light SM
neutrinos by combining
naturally suppressed Dirac and Majorana mass
scales together in a low-scale or ``mini'' seesaw mechanism.
Motivated by the AdS/CFT
correspondence~\cite{Maldacena:1997re}, we
consider a warped
extra dimension~\cite{Randall:1999ee} with a sub-TeV infrared (IR)
scale~\cite{Gripaios:2006dc,McDonald:2010iq,McDonald:2010fe}, in which the right-handed
neutrinos propagate. Via the
application of AdS/CFT to Randall-Sundrum (RS)
models~\cite{ArkaniHamed:2000ds}, this 5D model is dual to a 4D theory
with a hidden strongly
coupled sector of which the right-handed neutrinos are the
lightest fermionic composites. The SM fields, taken localized on the UV
brane, are fundamental objects in the sense that they are not part of
the hidden CFT. We work with an effective theory for energies
$\lesssim$~TeV, and depending on the UV completion
the SM fields may remain as fundamental or may themselves be
composites of a separate strongly coupled sector. 

Before proceeding we note that bulk gauge-singlet neutrinos
in RS models were considered in~\cite{Grossman:1999ra} and bulk SM
fermions in~\cite{Gherghetta:2000qt}. Subsequent studies of neutrino
mass appeared in~\cite{Huber:2002gp} and for
an incomplete list of recent works in this active field
see~\cite{Gherghetta:2003he,Gherghetta:2007au}. Related work on
right-handed neutrinos within a strongly coupled CFT was undertaken in
Ref.~\cite{vonGersdorff:2008is}. Our 
implementation within a sub-TeV scale effective theory
differs from these previous works. We also note that an order GeV  hidden
  sector has been invoked in connection with some recent
  experimental anomalies~\cite{Pospelov:2007mp,ArkaniHamed:2008qn}. A
  GeV scale mediator
 may be motivated by the leptonic cosmic ray
 anomalies\footnote{Our approach may also
  be of interest for models of TeV scale dark matter with a warped GeV scale
  mediator~\cite{BvH_KM}.}~\cite{ArkaniHamed:2008qn}, or 
  the dark matter may itself be $\sim$~GeV (see,
  e.g.,~\cite{Andreas:2010dz}). It is 
  sensible to ask what 
  consequences a sub-weak hidden sector may
  have in the neutrino sector. If the hidden sector contains fermions they
  could clearly influence the mechanism of neutrino mass
  generation. We present a specific framework here, but this matter
  may be of more general interest.
\section{Light Neutrinos from a Mini-Seesaw}
We consider a truncated RS model with a warped extra dimension described by the
coordinate $z\in[k^{-1},R]$. A UV brane of characteristic
energy scale $ k$ is located at
$z=k^{-1}$ and an IR brane with characteristic scale
$R^{-1}$ is located at $z=R$. The metric is given by
\beq
ds^2 = \frac{1}{(kz)^2}(\eta_{\mu\nu}dx^{\mu}dx^{\nu} -
dz^2)= G_{MN} dx^{M}dx^{N},
\label{bulkmetric}
\eeq
where $M,N,..$ ($\mu,\nu,..$) are the 5D (4D) Lorentz indices and
$k$ is the $AdS_5$ curvature. The characteristic IR scale is suppressed
relative to the curvature due to the warping, 
as is readily seen using the 
proper coordinate for the extra dimension.\footnote{This is defined by 
$y= k^{-1}\log(kz)$, where $y_i\in[0,L]$ and 
$L=k^{-1}\log(kR)$, in terms of which the IR scale is
exponentially suppressed, $R^{-1}=e^{-kL}k \ll k$.}
When sourced by a bulk cosmological constant and appropriate brane 
tensions the metric of Eq.~\eqref{bulkmetric}
is a solution to the 5D Einstein
equations~\cite{Randall:1999ee}. The length of the space is readily
stabilized~\cite{Goldberger:1999uk}. 

\begin{figure}[ttt]
\begin{center}
        \includegraphics[width = 0.3\textwidth]{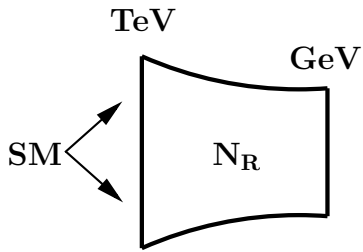}
\end{center}
\caption{Sketch of the ``Little Warped Space". The sterile neutrino $N_R$ propagates in a hidden warped space with IR scale of order
  a GeV and UV scale of order a TeV. The standard model
  resides on the UV brane and has a suppressed coupling to the chiral zero
  mode neutrino, which is localized toward the IR brane.}
\label{fig:nu_lit_throat}
\end{figure}

We take the SM to be localized on
the UV brane where the natural mass 
scale is $\sim k$, and accordingly take
$k\sim$~TeV. In addition to the SM we consider three singlet fermions
propagating in the bulk. We label these by $N_R$ as the
zero modes will be right-chiral fields that we identify as gauge-singlet
neutrinos. The IR scale is nominally taken to
be
$R^{-1}\sim$~GeV, but smaller values may be possible and will
be considered below.\footnote{Light  sterile
  neutrinos may be of interest in relation to the observed pulsar
  velocities~\cite{Fuller:2003gy} or as dark matter candidates (see e.g.~\cite{Kusenko:2010ik}).}  A sketch of the setup is given in
Figure~\ref{fig:nu_lit_throat}.  In analogy with the Little RS
model~\cite{Davoudiasl:2008hx} we can refer to this as a ``Little
Warped Space" (LWS) (for additional work on a truncated slice of $AdS_5$ see~\cite{Fukuyama:2007ph}). Such a truncated spacetime may seem
somewhat unusual at 
first sight. However, the setup can be thought of as an 
effective theory that describes the sub-TeV scale physics of a more
complete theory, enabling one to consider the effects of a light
warped/composite hidden sector without having to specify the supra-TeV
physics. In general the supra-TeV effects will be encoded in
   UV localized effective operators. As with any sub-TeV scale
   effective theory, the effective operators that break
   approximate/exact symmetries of the SM must be
   adequately suppressed  to
   ensure that problematic effects like rapid p-decay and excessive flavour
   violation do not occur.

To demonstrate our main points it will suffice to consider a single
generation of neutrinos. The action for a bulk fermion $N_R$ in the
background (\ref{bulkmetric})  is
\begin{eqnarray}
S_{\nl}&=&\int
d^5x\sqrt{G}\left\{\frac{i}{2}\bar{N}_R\Gamma^{B}e^A_{B}\partial_A
  N_R-\frac{i}{2}(\partial_A\bar{N}_R)\Gamma^{B}e^A_{B}N_R+ck\bar{N}_RN_R\right\},
\end{eqnarray}
where $\Gamma^{\mu,5}=\{\gamma^\mu,i\gamma^5\}$ are the 5D Dirac-gamma
matrices, $e^A_{B}$
is the f\"unfbein and we write the Dirac mass in units of the
curvature $k$. We have dropped the spin-connection terms which
cancel in the above. A Kaluza-Klein (KK) decomposition may be
performed as
\bea
N_R(x,z)=(kz)^2\sum_n \left\{\nu_L^{(n)}(x)f_L^{(n)}(z)+\nu_R^{(n)}(x)f_R^{(n)}(z)\right\},
\eea
and the bulk wavefunctions $f^{(n)}_{L,R}$ for the chiral components
of $N_R$ can be readily found~\cite{Grossman:1999ra}. The boundary
conditions force one chirality to be odd and without loss of
generality we take this
to be the left-chiral field. The single massless mode in the
spectrum then has right-chirality. The KK-expanded Lagrangian is
\begin{eqnarray}
S_{\nl}&=&\sum_{n}\int
d^4x\left\{\bar{\nu}^{(n)}\gamma^{\mu}\partial_\mu
  \nu^{(n)}-m_n\bar{\nu}^{(n)}\nu^{(n)}\right\},
\end{eqnarray}
where $\nu^{(n)}=\nu_L^{(n)}+\nu_R^{(n)}$ is a Dirac fermion with KK
mass $m_n$  for
$n>0$ and $\nu^{(0)}=\nu^{(0)}_R$ is the massless right-chiral zero
mode. Its bulk profile is
\bea
f^{(0)}_R(z)=\sqrt{\frac{k(1+2c)}{(kR)^{1+2c}-1}}\left(kz\right)^{c},
\eea
where the dimensionless mass parameter $c$ controls its localization
along the extra dimension. We will be interested in IR localization
with $c\simeq1$ as this reduces the wavefunction overlap of the zero
mode with UV localized SM neutrinos and therefore suppresses
the Dirac mass below the weak scale.

At this point there is no lepton number violation and the spectrum
consists of a single Weyl neutrino and a tower of Dirac neutrinos with
masses $m_n\sim n\pi/R$. We can introduce
lepton number violation in the form of a marginal
operator on the IR brane:
\begin{eqnarray}
S_{\nl}&\rightarrow&S_{\nl} -\frac{\lambda_{\nl}}{2}\int
d^5x\sqrt{-g_{ir}}\left\{\bar{N}_R^cN_R
  +\mathrm{H.c.}\right\}\delta(z-R)\nonumber\\
&=&\sum_{m,n}\int
d^4x\left\{\bar{\nu}^{(n)}\gamma^{\mu}\partial_\mu
  \nu^{(n)}-m_n\bar{\nu}^{(n)}\nu^{(n)}-\frac{M_{mn}}{2}(\bar{\nu}_R^{(m)})^c\nu^{(n)}_R+\mathrm{H.c.} \right\},
\eea  
where the effective Majorana masses are
\bea
M_{mn}&=& \left. \lambda_{\nl}f_R^{(m)}f_R^{(n)} \right|_{z=R}.
\eea
Note that $\nu_L^{(n)}$ does not acquire a boundary Majorana mass as $\left. f_L^{(n)} \right|_{R}=0$. For IR localization of interest to us the zero mode Majorana mass takes a particularly simple form:
\bea
M_{00}&\simeq& (1+2c)\frac{\lambda_{\nl}}{R}.\label{maj_zero_mode}
\eea
Using the results in the
Appendix for $c\simeq 1$, the Majorana masses for the $m,n>0$
modes can be approximately related to that of the zero mode:
\bea
|M_{0n}|\simeq \sqrt{\frac{2}{2c+1}}M_{00}\quad\mathrm{and}\quad |M_{mn}|\simeq
\frac{2M_{00}}{(2c+1)}\quad\mathrm{for}\quad m,n>0\ .\label{maj_higher_mode}
\eea
We note that $M_{mn}\sim \lambda_{\nl}/R$ for all $m,n$, as expected for an IR
localized mass.
These Majorana masses mix\footnote{One could
instead include the boundary mass in
  the IR boundary conditions and obtain the full KK spectrum
  directly~\cite{Huber:2002gp}. However our main points are easily
  seen treating the
  boundary terms as perturbations.}  the KK modes and the true mass
eigenstates are linear combinations of  $\nu^{(n)}$. The spectrum
consists of a tower of Majorana neutrinos with masses starting at $\sim R^{-1}$.
For $\lambda_{\nl}\sim 0.1$ one has $M_{00}\sim  R^{-1}/10$ and the
lightest mode is predominantly composed
of $\nu^{(0)}_R$. The higher modes are pseudo-Dirac neutrinos with
mass splittings set by $M_{mn}< m_n$. The Dirac mass $m_n$ increases with $n$ as
$m_n\sim (n+c/2)\pi/R$ while $M_{mn}$ does not
significantly change, so the Majorana masses become increasingly
unimportant for the higher modes.

Having determined the spectrum of sterile neutrinos we can proceed to
consider their coupling to the SM. This occurs via a UV localized
Yukawa interaction
\bea
S&\supset &-\frac{\lambda}{\sqrt{M_*}} \int d^5x\sqrt{-g_{uv}} \ \bar{L}
H N_{R}\ \delta(z-k^{-1})\ ,
\eea
where $L$ is a lepton doublet and $H$ is the SM scalar doublet. After
integrating out the extra dimension this generates Dirac mass terms
coupling the KK neutrinos to the SM:
\bea
S&\supset &-\sum_n  \int d^4x\ m^{\D}_n\  \bar{\nu}_L \nu^{(n)}_{R}\  ,
\eea
where $m^{\D}_n  = \lambda \langle H\rangle
f^{(n)}_R(k^{-1})/\sqrt{M_*}$ and $\langle H\rangle\simeq 174$~GeV is the vacuum
value of the SM scalar. For the zero mode this gives
\bea
m^{\D}_0 &\simeq&  \lambda\ \sqrt{\frac{k(1+2c)}{M_*}} \ (kR)^{-c-1/2}\ \langle H\rangle\ ,
\eea
and the results from the Appendix can be used to find $m^{\D}_n$ for $n>0$. 

Including the boundary coupling to SM neutrinos produces a somewhat
complicated mass matrix describing both the SM and KK neutrinos
(similar to that in~\cite{Huber:2002gp}). Despite this the
hierarchy of scales generated by the KK profiles allows the basic
spectrum to be readily understood. For $n>0$ the previously
pseudo-Dirac neutrinos now have Dirac masses coupling them to the SM,
which are given by $(m^{\D}_n/m_n)\simeq \lambda
(\langle H\rangle/k) \sqrt{2/M_*R}\ll 1$ for $c\simeq1$.
This coupling can essentially be
neglected to leading order and therefore the $n>0$ modes remain as
pseudo-Dirac neutrinos comprised of predominantly sterile KK modes.  

The coupling of the SM to the zero mode is more important  as this mode is
not strongly coupled to a KK partner. To leading order the zero mode
and the SM neutrino essentially form a standard seesaw pair. The heavy
mode has mass $\simeq M_{00}$ and is mostly comprised of $\nu^{(0)}$,
while the mass of the light SM neutrino is of the usual seesaw
form,
\bea
m_\nu\  \simeq\ \frac{(m^{\D}_0)^2}{M_{00}}\ \simeq\ \frac{\lambda^2}{\lambda_{\nl}}\ \frac{\langle
  H\rangle^2}{M_*}\ (kR)^{-2c}\ .
\eea 
To get a feeling for the scales involved we plot the inverse radius $R^{-1}$ as a function of $c$
for fixed values of $m_\nu=1$~eV and $m_\nu=10^{-2}$~eV in
Figure~\ref{fig:radius_vs_c}. The following set of parameter
values is used for the plot: $k=1.5$~TeV,
$\lambda/\sqrt{2}= \lambda_{\nl}=0.1$ and $k/M_*=1/6$. We also restrict $|c|$ to be
no larger than the values employed in~\cite{Davoudiasl:2008hx}. It is
clear that the SM neutrino mass is readily suppressed below the weak
scale to within the range of interest for the solar and atmospheric
neutrino data. 

We observe that this  approach generates naturally suppressed neutrino
masses in a two-fold process. Firstly, the effective 4D Dirac
and Majorana mass scales are suppressed; the sub-TeV Majorana
scale ($M_{00}\ll m_W$) is generated by warping while the Dirac mass
is suppressed by
a small wavefunction overlap ($m^{\D}_0\ll M_{00}\ll m_W$). Secondly, a  low-scale or ``mini'' seesaw mechanism operates
between the lightest KK mode and the SM neutrino, serving to further
suppress the SM neutrino mass. Together these
elements realize the order eV neutrino masses. We emphasize that,
unlike most seesaw models with light
sterile Majorana mass scales, the small 4D masses are generated naturally
and do not require tiny couplings.  

Note that Figure~\ref{fig:radius_vs_c} includes values of $R^{-1}$ much less
than a GeV, showing that naturally light neutrino masses can be
generated for a range of IR scales. Clearly, the phenomenological
consequences of generating neutrino mass in this way depend on the
specific value of $R^{-1}$, and one cannot make general statements
over such a wide range of energies. A detailed study would be required
to determine the relevant limits on $R$,
which is beyond the scope of this work. However, we will offer some
comments on aspects of the hidden sector for parts of
this range in Section~\ref{spectrum}, focusing on the case of
$R^{-1}\sim$~GeV. We emphasize that while light neutrino masses can be
obtained with $R^{-1}\ll$~GeV, models with light values of $R^{-1}$ may face
severe bounds from experimental data. These require further
detailed investigation to determine their viability.
\begin{figure}[t] 
\centering
\includegraphics[width=0.5\textwidth]{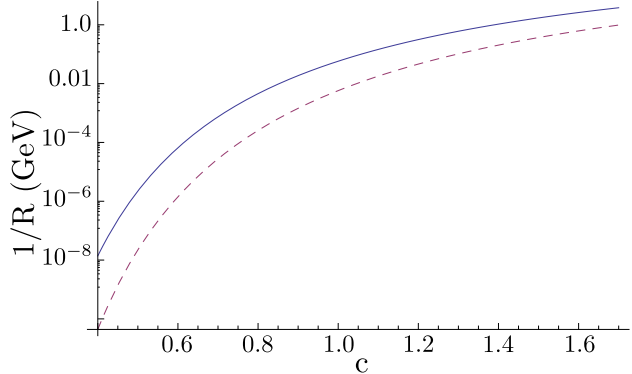} 
  \caption{Plot of the inverse radius $R^{-1}$ as a function of the
    bulk mass parameter $c$ for fixed values of the light neutrino
    mass $m_\nu$. We fix $m_\nu= 1$~eV ($m_\nu= 10^{-2}$~eV) for the
    solid (dashed) curve. The radius is related to the mass of the lightest right-handed
    neutrino as $M_{00}\sim R^{-1}$.}
  \label{fig:radius_vs_c}
\end{figure}

\section{4D Gravity and AdS/CFT}
The particle spectrum will contain
additional sub-TeV modes in the form of metric fluctuations. These are localized
towards the IR and their precise coupling to the UV localized SM fields depends
on the UV action~\cite{Davoudiasl:2003zt}. If one seeks to include 4D gravity then a
UV localized Einstein-Hilbert term with strength $M_{UV}$
will appear to reproduce the 4D Planck mass as $M_{Pl}^2\sim M_{UV}^2 +M_*^3/k$.
Here $M_*^3/k$ is the usual bulk contribution to the 4D
Planck mass in RS models but is suppressed relative to the 4D Planck scale
for the LWS. Including this boundary term is akin to the 4D approach
of retaining a $\sim$~TeV scale cutoff for the SM while including Einstein gravity. For
$M_{UV}\rightarrow\infty$ gravity 
decouples and the KK gravitons acquire 
a Dirichlet UV boundary condition as in
Ref.~\cite{Davoudiasl:2008hx}. In this case
they do not directly couple to the SM. Alternatively, 4D gravity could be
included by lowering the fundamental
gravity scale
   to $\sim$~TeV. This requires large
    extra dimensions transverse to the warping but has the advantage
    of potentially solving the SM hierarchy
   problem~\cite{ArkaniHamed:1998rs}. The 4D Planck mass would then be
   given by $M_{Pl}^2 \sim M_*^{3+d}V_d /k$, where $V_d$ is the volume
   of the $d$-dimensional transverse space. We focus on the case of a
   large UV boundary term in the remainder of this work.

It is helpful to comment on the dual 4D description of the LWS, obtained via the
application of the AdS/CFT
correspondence to RS models~\cite{ArkaniHamed:2000ds}, before
discussing the hidden-sector particle spectrum further. The present
model would
be considered dual to a 4D theory in which the UV localized SM fields are treated as
fundamental objects external to a hidden strongly coupled sector. The
composite objects correspond to IR localized fields in the 5D picture
(roughly; see e.g.~\cite{Batell:2007jv}) and the lightest fermionic composite is the right-handed neutrino (or
neutrinos in a three generation model). Note that the IR localization
of the Majorana mass term
indicates that lepton number is broken within the strongly coupled
sector, distinct to~\cite{Gherghetta:2003he} in which lepton number is
broken in the UV.  

A discussion of the modified
application of AdS/CFT for truncated warped spaces is given
in~\cite{Davoudiasl:2008hx} and similar points remain valid in the
present work. The dual theory is conformal for energies
$\mathrm{TeV} \gtrsim E\gtrsim R^{-1}$, with the conformal symmetry broken
spontaneously in the IR (dual to the IR brane) and explicitly in the
UV (dual to the TeV brane, itself ascociated with the scale of electroweak symmetry
breaking). For every bulk field in the 5D picture there exists a CFT
operator in the dual 4D theory that is sourced by a
fundamental field. The source field corresponds to
the UV-brane value of the given bulk field in the 5D theory. If the
 bulk field does
not possess a UV-localized brane kinetic term, the source field has
\emph{no bare kinetic term} in the 4D theory and is seemingly non-dynamical. However, due to the coupling to the CFT, an
\emph{induced} kinetic term arises, so the ``source'' is
in fact dynamical~\cite{Agashe:2002jx}.\footnote{Strictly speaking, a brane
kinetic term is necessary to regulate the 5D theory~\cite{Georgi:2000ks}, so this dicussion
corresponds to the case where, after divergences are removed, the
contribution to the renormalized kinetic term from the bare kinetic term is
dominated by the CFT-induced kinetic term.} 

The dual description can shed light on the above discussion of the 4D Planck mass. The dual 4D theory for RS1
corresponds to a CFT plus 4D Einstein gravity~\cite{ArkaniHamed:2000ds}.  In the absence of a UV
localized Einstein-Hilbert term, the dual 4D theory does not contain a
bare Einstein-Hilbert term. However, the coupling of the 4D graviton to
the CFT induces an Einstein-Hilbert term so that gravity
acquires its usual Einstein dynamics. Thus the
4D Planck mass obtained in RS1, namely $M_{Pl}^2\simeq M_*^3/k$, is
predominantly induced by the CFT~\cite{ArkaniHamed:2000ds}. Adding a
UV localized kinetic term is dual to including a bare Einstein-Hilbert
term in the 4D theory. The 4D Planck mass then becomes $M_{Pl}^2\simeq M_{UV}^2+
M_*^3/k$, where the first (second) term is a bare (CFT-induced)
contribution. Thus the 4D Planck mass is no longer completely
induced by the CFT. In the RS model the CFT induced part of the Planck
mass is 
typically taken to dominate (or is the same order as) the UV piece, so the
CFT-induced part of the Planck mass plays an essential role in determining the
dynamics of the 4D graviton.

One can consider a 4D CFT, however, with a UV cutoff
much less than the Planck scale, that does not completely induce the
Planck mass. In the 5D picture
this corresponds to a slice of $AdS_5$ with a UV cutoff much less than
the Planck scale, $M_*\ll M_{Pl}$. This is the case considered in the
Little RS model~\cite{Davoudiasl:2008hx}, where the UV cutoff is of order
$10^3$~TeV.  In Ref.~\cite{Davoudiasl:2008hx} the bulk graviton was
given a Dirichlet UV BC, thereby removing the zero-mode graviton and projecting
4D gravity out of the
theory. The Dirichlet UV BC can be realized by sending the
coefficient of the UV
Einstein-Hilbert term to infinity,
$M_{UV}\rightarrow\infty$.\footnote{See Ref.~\cite{Davoudiasl:2003zt}
  for a detailed discussion of the modified UV BC for gravity with
  brane localized curvature.} In the dual picture Poincar\'e symmetry
is no longer a gauged dynamical
symmetry but rather a global symmetry. Though not essential for the TeV-scale
physics of interest in Ref.~\cite{Davoudiasl:2008hx},
to be a completely realistic low-energy theory the Little RS model
should of course include 4D Einstein gravity. One
can ask how 4D gravity can be included in the model. The
answer is to simply include a large (but finite) UV localized
Einstein-Hilbert term such that $M_{Pl}^2\simeq M_{UV}^2\gg
M_*^3/k$. In this case the CFT-induced part of the Planck mass is
subdominant and the 4D Planck mass is essentially an input
parameter, separate from the scale at which the CFT behaviour breaks down. The particle physics cut-off remains at $M_*\ll
M_{Pl}$ and the theory requires some UV completion at energies
$E>M_*$, but the low-energy theory now includes the usual coupling to 
4D Einstein gravity in addition to the (broken) CFT.

A similar discussion carries over for the dual description of the
LWS. Whether or not the dual theory contains Einstein
gravity depends on the UV action. With an infinite UV localized Einstein-Hilbert
term, the 4D
graviton is projected out and the dual theory is purely that of
fundamental SM fields and a
hidden CFT, both possessing a global Poincar\'e symmetry. Einstein
gravity is included in the dual 4D
theory by retaining a finite value for $M_{UV}$ in the 5D theory. Provided
$M_{UV}^2\simeq M_{Pl}^2\gg M_*^3/k$ the Planck mass is an input parameter whose
origin is separate from the CFT dynamics (though the CFT
induces a subdominant contribution of order $M_*^3/k$). Note that the
UV cutoff for the
theory remains at the TeV scale, $M_*\sim$~TeV, but the low-energy
theory also includes 4D Einstein gravity. This is akin
to cutting off the SM (or any other 4D theory) at the TeV scale
while retaining 4D gravity, despite the fact that
$M_{Pl}\gg$~TeV. 

We now turn to the dual description of the bulk neutrino. In the
absence of the IR localized Majorana mass term, the dual 4D
theory for  $c> 1/2$ is given
schematically by the Lagrangian~\cite{Contino:2004vy}
\bea
\mathcal{L}_{4D}^N\  \sim\  \mathcal{L}_{CFT} +g_N\  k^{1/2-c} \psi_R\
\mathcal{O}_{N}\ +\ .\ .\ .\ \ .\label{eq:cft_N}
\eea
Here $\psi_R$ is a fundamental source whose dynamics are induced by
the CFT (absent a UV localized kinetic term), $g_N$ is a dimensionless coupling, and the fermionic composite
CFT operator $\mathcal{O}_N$ has
dimension $3/2+|c+1/2|$. Observe that the mixing operator
between the fundamental field and the CFT operator in
Eq.~(\ref{eq:cft_N}) is irrelevant. Pulling a factor of
$\mu^{1/2+c}$ out of $\mathcal{O}_N$ to give the fermionic operator a canonical
dimension at the scale $\mu$, we see that the mixing becomes
tiny in the IR, $\mu\ll k$. The source $\psi_R$ is
determined by the UV value of the bulk field $N_R$, and the
relationship between $\psi_R$  and the chiral mode depends on the
localization parameter $c$. For the values of interest here, the
zero-mode is localized towards the IR brane and has very little overlap
with the UV brane. Therefore the source contains only a tiny admixture of
the zero-mode and the physical chiral-mode corresponds predominantly to
a composite CFT state. The tower
of Dirac fermions consists mostly of composites but contains an
admixture of $\psi_R$.

The small Yukawa coupling between the SM and the lightest right-chiral neutrino
is also understood in the dual picture. The SM is external to the CFT,
but couples directly to the source field $\psi_R$. Writing this field
in terms of the physical 
mass eigenstates introduces the aforementioned tiny mixing
angle, the dependence on which ensures the effective Yukawa couping is highly
suppressed. Turning on the IR Majorana mass gives the chiral mode a
 mass and splits the Dirac fermions. This IR term only affects
CFT correlation
functions at distances larger than the conformal length of the space,
$\Delta x\gtrsim R$. The Majorana mass should therefore be on the
order of $\sim R^{-1}$, in agreement with Eqs.~(\ref{maj_zero_mode})
and (\ref{maj_higher_mode}). 

\section{Comments on the Hidden Sector Spectrum\label{spectrum}}

The hidden sector will contain a tower of KK gravitons with mass
splittings of order GeV (or less for 
smaller $R^{-1}$). On the surface it might seem that the light mass of these
KK gravitons could
be phenomenologically troublesome. However, as mentioned above, the
strength with which these modes couple to the SM depends on the UV
action. Consider the case of a large UV localized Planck mass
$M_{Pl}\simeq M_{UV}$. Then, by construction,  the UV values of the
graviton wavefunctions are highly suppressed, such
that the coupling between the UV localized SM and the KK gravitons is of
order $M_{Pl}^{-1}$. Much like the KK gravitons with masses below the
weak scale in RS2~\cite{Randall:1999vf}, the KK gravitons are not
phenomenologically worrisome as their direct coupling to the SM simply
produces subdominant corrections to the Newtonian potential.
Indeed,
for $M_{Pl}\simeq M_{UV}$  the coupling of KK gravitons to the SM will
essentially match those in Ref.~\cite{McDonald:2010fe}, where a
detailed analysis of a warped hidden sector containing order GeV KK
gravitons found them to be viable. The similarity to RS2 makes
obvious the fact that even sub-GeV IR scales are viable as far as the
KK gravitons are concerned. In the limit
$M_{UV}\rightarrow\infty$ the graviton wavefunctions will be banished from
the UV brane and the KK gravitons have no direct coupling to the
SM.\footnote{Note that taking $M_{UV}\rightarrow\infty$ is not the
  same as taking the UV cutoff to infinity; the UV cutoff for the
  truncated warped space ($M_*$) remains at the TeV scale throughout this
  discussion. $M_{UV}$ refers only to the coefficient of the UV-localized
  Einstein-Hilbert term. The use of large boundary terms to interpolate between mixed and
  Dirichlet BCs in models on an interval is discussed in,
  e.g.,~\cite{Csaki:2005vy}. In our case the large UV term provides a
  convenient way to
  interpolate between a finite 4D Planck scale and a Dirichlet UV BC
  (or, equivalently, an infinite 4D Planck mass).} Then the model is
purely a particle physics framework, involving no scales beyond the
TeV scale, and does not include low-energy gravity effects (as in
Ref.~\cite{Davoudiasl:2008hx}). Note that one cannot consider a purely
Neumann UV BC for the graviton as then the 4D graviton would couple
with strength $M_*^3/k\sim$~TeV. The UV BC must be either
Dirichlet or include the effects of a large UV term. 

Let us emphasize that, even with a large UV
Einstein-Hilbert term, the particle physics cutoff on the warped space
remains at the TeV scale, in connection with the scale of electroweak
symmetry breaking. The most general UV action will contain the usual series of
non-renormalizable operators involving SM fields, which encode the details
of the UV completion. These operators can induce
unwanted effects (like rapid p-decay) and experimental bounds  require the dimensionless
coefficients of such operators to be sufficiently small. This is true
of any sub-TeV scale effective theory. The inclusion of a large UV
Einstein-Hilbert term is not inconsistent with retaining a TeV-scale
UV cutoff as this is to be viewed purely as a phenomenological tool
with which to include the low-energy effects of gravity in the
theory. The cutoff remains at the TeV scale, where new physics
associated with, e.g., stabilization of the weak scale, should
appear. As mentioned earlier, this approach, motivated by AdS/CFT considerations, is akin to including
low-energy gravity effects in particle physics models like the SM,
even though new physics is likely to appear well before the Planck
scale to solve, e.g., the hierarchy problem. The key particle physics
point we seek to make in this work, regarding a sub-TeV seesaw
mechanism, does not require the inclusion of low-energy 4D gravity. We
have discussed a way to include 4D gravity in the theory for
completeness, but
 as far as the neutrino sector is concerned, our key points are seen
simply by taking a Dirichlet UV BC for gravity and decoupling the 4D
graviton. In this case the KK gravitons do not couple directly to the
SM.

Independent of their coupling strength to the SM the
KK gravitons
possess sizable couplings to the KK neutrinos. The KK neutrinos and
gravitons are both localized towards the IR brane and in general have
large wavefunction overlaps, so the relevant coupling strength is of
order $R^{-1}\gg M_{Pl}^{-1}$.  This coupling allows a given KK
graviton $h_a$ to decay to lighter KK neutrinos $h_a\rightarrow 
\nu^{(m)}\nu^{(n)}$, with width $\Gamma_a\sim (k/M_*)^3 m_a^3 R^2$
(similar to the case for IR localized fields~\cite{Davoudiasl:1999jd};
also see Ref.~\cite{McDonald:2010fe}).
Absent hierarchically small values of $k/M_*$ these decays will be
prompt for $R^{-1}\sim$~GeV. The radion $r$ will also decay via
$r\rightarrow
\nu^{(0)}\nu^{(0)}$, provided $m_r$ adequately exceeds the zero mode mass,
$m_r\gtrsim 2M_{00}$, with a typical coupling strength set by the IR
scale $R^{-1}$.

The metric fluctuations play an important role in the phenomenology of the KK neutrinos. The $n>0$ KK
neutrinos will promptly decay via graviton production,
$\nu^{(n)}\rightarrow h_a \ \nu^{(m)}$. The widths go like
$\Gamma_n\sim (k/M_*)^3 m_n^3R^2$ and are increasingly broad as one
goes up the KK tower. For a given value of $n$, decays with $n\sim
a+m$ are preferred. Decays to the SM are also possible but these have
to go through the Dirac Yukawa-coupling, which is necessarily
suppressed by a small wavefunction overlap to ensure light SM neutrino
masses. Therefore the SM decays are highly suppressed and, when
available, decays to lighter KK neutrinos will dominate. The
preference for decays with $n\sim a+m$ means production of a large-$n$
KK neutrino will create a cascade decay down the KK tower.

 On the other hand, the zero mode
neutrino can only decay to the SM and must therefore decay through the
Dirac Yukawa-coupling. These decays will be much slower than the hidden sector KK
decays. Clearly, if the light neutrinos are very long lived and are
thermalized in the early universe they could present cosmological
difficulties, depending on the IR scale. At energies above the IR
scale the RS geometry is
replaced by an AdS-Schwarschild space~\cite{Witten:1998zw}, with the
phase transition to the RS-phase occurring roughly at the IR scale,
though the precise behaviour is sensitive to the details of the
stabilization mechanism~\cite{Creminelli:2001th}. If the IR scale is
of order GeV, this phase transition would occur prior to
BBN.\footnote{Note that provided
the IR scale adequately exceeds about 4~MeV to allow BBN,
it is possible that primordial inflation never reheated beyond
the IR 
scale, in which case the details of the phase transition are
irrelevant.} In this case the
hidden spectrum would not contain any fields lighter than $\sim$~GeV,
except possibly the right-chiral neutrinos $\nu^{(0)}$,
depending on the value of the localization parameter $c$. Order GeV
sterile neutrinos are a common feature of, e.g., the
$\nu$MSM~\cite{Gorbunov:2007ak}, and
interesting scenarios are known to be viable. The decay of sterile neutrinos can
produce interesting consequences and, in particular, admit
non-standard leptogenesis scenarios~\cite{Gorbunov:2007ak}.

For IR scales much less than $\mathcal{O}(\mathrm{MeV})$ the phase
transition would occur after BBN and there would be many light degrees
of freedom in the spectrum. An important point to note is that any
interactions between the UV localized SM fields and the IR localized
KK/CFT-modes must proceed through the propagator of a bulk
field. The propagating field can be either a bulk neutrino or the
graviton, but the tiny coupling of the latter to the SM suppresses the
graviton effects. A key point is that the UV-to-IR propagator for a
bulk field on a slice of $AdS_5$ essentially cuts off for distances
$z^{-1}\gtrsim E$ when the injection energy $E$ greatly exceeds the IR
scale, $E\gg R^{-1}$~\cite{ArkaniHamed:2000ds}. Interactions of the
CFT that could potentially bring the the light KK modes into thermal
equilibrium for $R^{-1}\ll $~MeV therefore turn out to be unimportant
at energies much larger than the IR scale, $E\gg R^{-1}$, as would be
relevant for BBN. Thus the light KK modes need not be brought into
equilibrium prior to BBN by CFT interactions.

Oscillations between light right-handed neutrinos and SM neutrinos, on
the other hand, have the potential to bring the lighter modes into
equilibrium. A detailed numerical analysis including the effects of
mixing, and the tower of thresholds, would be required to determine
the extent to which equilibrium is reached. We note that composite
right-handed neutrinos with very light seesaw scales can imprint
signals in the CMB which can be in conflict with the
data~\cite{Okui:2004xn}, though the case of a hidden CFT with
low-scale neutrino composites has not been studied in detail. With
very light IR scales one could abandon the seesaw and simply consider
light Dirac neutrinos, as in Ref.~\cite{Okui:2004xn}. The neutrino mass would
then be
$m_\nu\equiv m_0^D\sim \langle H\rangle/(kR)^{c+1/2}$, and taking,
e.g., $R^{-1}\sim
10$~eV and $c\simeq 0.8$ gives $m_\nu\sim 0.1$~eV, which is in the
interesting range.
\section{Conclusion}
We have developed a mini-seesaw mechanism in which light
neutrino masses are achieved by combining naturally suppressed
Dirac and (sterile) Majorana mass scales together in a low-scale seesaw
mechanism. The model is motivated by the AdS/CFT
correspondence~\cite{Maldacena:1997re} and the
notion of light composite right-handed
neutrinos~\cite{ArkaniHamed:1998pf}, and is dual to a 4D theory with a
strongly coupled hidden sector whose lightest fermionic composites are
right-handed neutrinos. The model employs a truncated (``little'') warped space 
that is dual to a 4D theory possessing conformal symmetry in some window, $M_*> E>R^{-1}$
with $M_*\ll M_{Pl}$. Depending
on the UV completion
the theory may or may not be conformal at higher energies $E>M_*$. It would be interesting to investigate these
ideas further
to consider the viability of full three-family models and perform a
detailed study of the bounds on, and
phenomenology of, the light KK neutrinos. As a speculation, it may be possible
to construct
theories of flavor using these ideas, either by considering flavor
structures for sterile neutrinos within
the LWS or along the lines
of~\cite{Abel:2010kw}. We
will investigate some of
these matters in a future work.

The author thanks E.~ Akhmedov, T.~Frossard, W.~Rodejohann, R.~Takahashi and J.~Ward.
\appendix
\section*{Appendix}
The bulk profiles for the $n>0$ KK fermions are~\cite{Grossman:1999ra}
\begin{eqnarray}
f_{L}^{(n)}(z)&=&-\frac{\sqrt{kz}}{N_{n}}\left\{J_{\alpha_{\li}}(m_{n}z)+\beta_{n}Y_{\alpha_{L}}(m_{n}z)\right\},\\
f_{R}^{(n)}(z)&=&\frac{\sqrt{kz}}{N_{n}}\left\{J_{\alpha_{\ri}}(m_{n}z)+\beta_{n}Y_{\alpha_{R}}(m_{n}z)\right\},
\end{eqnarray} 
with the order of the Bessel functions being
$\alpha_{\li,\ri}=|c\pm1/2|$ with $c$ defined in the text. The KK masses are determined by
$J_{\alpha_{\li}}(m_n R)\simeq0$, giving $m_n\simeq(n+c/2)\pi R^{-1}$ for
$n>0$. For $c\simeq 1$ these masses are $m_nR\simeq 4.6\ ,\ 7.7\ ,\ 10.8\ ...$\ .
When $f_{L}^{(n)}(z)$ has Dirichlet BCs the $n>0$ right-chiral modes take the following
boundary values for $c\simeq1$ and $m_n<k$:
\bea
f_{R}^{(n)}(k^{-1})&\simeq&\frac{1}{\Gamma(c+1/2)}\sqrt{\frac{2\pi}{R}}\left(\frac{m_n}{2k}\right)^c\
,\nonumber\\
f_{R}^{(n)}(R)&\simeq&(-1)^n\sqrt{\frac{2}{R}}\ .
\eea



\end{document}